\documentclass[aps,pra,10pt,english,showpacs,floatfix,twocolumn,twoside]{revtex4-1}

\usepackage{amsmath,amstext,amssymb,babel,graphicx,geometry,bm,dcolumn,color,hyperref,times}
\begin{document}

\title{
To freeze or not to: Quantum correlations under local decoherence  
} 
\author{Titas Chanda,  Amit Kumar Pal, Anindya Biswas, Aditi Sen(De), and Ujjwal Sen }
\affiliation{Harish-Chandra Research Institute, Chhatnag Road, Jhunsi, Allahabad - 211019, India }

\begin{abstract}
We provide necessary and sufficient 
conditions for freezing of quantum correlations as measured by quantum discord 
and quantum work deficit in the case of bipartite as well as multipartite states 
subjected to local noisy channels. We recognize that inhomogeneity of the magnetizations
of the shared quantum states plays an important role in the freezing phenomena. 
We show that the frozen value of the quantum correlation and the time interval for freezing 
follow a complementarity relation. 
For states which do not exhibit ``exact'' freezing, but can be frozen ``effectively'', by having a very slow 
decay rate with suitable tuning of the  state parameters,
we introduce an index -- the freezing index -- to quantify the goodness of freezing.
We find that the freezing index can be 
used to detect quantum phase transitions and discuss the corresponding scaling behavior.
\end{abstract}


\maketitle

\section{Introduction}
\label{intro}

Characterizing correlations between different 
subsystems of a composite quantum system has been an important field of research in 
quantum information~\cite{hhhh,modi_rmp_2012}. 
This is due to the fact that quantum correlations, in the form of entanglement, is 
shown to be significantly more useful for performing communication and computational tasks 
over their classical counterparts~\cite{ref1}. Moreover, these tasks 
have successfully been realized in the laboratory in several physical systems and thereby attracted 
a lot of attention in the field of detection and quantification of quantum correlations
~\cite{exp}. On the other hand, several non-intuitive results like local 
indistinguishability of orthogonal product states, non-classical efficiencies of a 
computational task by states having negligible entanglement, etc. have also been 
discovered~\cite{ref2,ref3,dqci_group},
which highlights the needs for conceptualization of quantumness in a composite
system that is different from entanglement. This led to the introduction of
quantum correlations like quantum discord (QD)~\cite{disc_group} and 
quantum work deficit (QWD)~\cite{wdef_group} that are independent of the entanglement paradigm.

One of the difficulties encountered  in realizations of quntum information protocols is that quantum correlations decohere rapidly by 
interaction with the environment \cite{decohere_rev}. Specifically, the intra-system quantum correlations decrease with time 
while the quantum correlations between the system and the environment increase. As a result, the system 
typically becomes less efficient in performing quantum information processing tasks. 
Therefore, the decay of quantum correlations with time in an open 
quantum system is a cause for concern. 
Consequently, knowledge of the behavior of quantum correlations 
in various quantum systems when subjected to different environments seems indispensable.   
Recent studies show that for a specific class of states, entanglement undergoes a 
\textit{sudden death}~\cite{sd_group,sd_group2,sd_group3} at a finite time whereas QD decays asymptotically with 
time~\cite{disc_dyn_group,disc_dyn_group2,disc_dyn_group3,mazzola_prl_2010,adesso_pra_2013,disc_freez_group,disc_expt_group}.

Although a few studies have addressed the issue of preserving quantum correlation measures 
during their dynamics~\cite{mazzola_prl_2010,adesso_pra_2013,disc_freez_group,disc_expt_group}, 
identifying the inherent property in quantum states, prohibiting 
the loss of quantum correlations over time, is still an open question. In the present work, we derive 
 necessary and sufficient conditions for freezing of quantum correlation measures, for both QD and 
QWD, when a two-qubit state \emph{with magnetization} is 
subjected to local depolarizing channels, in which bit-flip (BF), phase-flip (PF), or 
bit-phase-flip (BPF) errors can occur. We show that inhomogeneity in magnetization plays a crucial role for the 
freezing behavior of the state. The necessary and sufficient criteria show that there exist  regions in which 
QD freezes while QWD does not, highlighting the necessity of proper choice of the quantum correlation measure
to demonstrate the freezing phenomenon, and  also that 
if the efficient performance of a certain quantum information task requires the freezing of a particular 
quantum correlation measure, the same may not remain efficient in a situation or environment where 
another quantum correlation is frozen.
For both QD and QWD, we propose a complementarity relation between the quantum correlation during the freezing 
interval and the duration of the interval.  For the class of states for which freezing is observed, we find a 
correspondence between the entanglement of the initial state and its freezing properties.  
The study is extended to multipartite states where two prescriptions for generating 
multipartite freezing states are proposed. Like decoherence-free subspaces \cite{decohere_free}, introduced to protect qubits, 
the generated inherently decoherence-free states can be a building block of quantum memory \cite{q_mem} in which quantum correlation
in the form of QD or QWD can be stored.

Undoubtedly, the bipartite as well as the multipartite states which show freezing are of immense theoretical and 
experimental importance 
in quantum information processing tasks as they are inherently 
decoherence-free states for a finite time interval. However, there exists a large class of states for which 
quantum correlations do not freeze, but the change in quantum correlations is very slow with time. Hence it is interesting 
to quantify the freezing quality of a quantum correlation measure. We introduce a measure to quantify the goodness 
of freezing and call the measure as the \textquotedblleft freezing index\textquotedblright. We then apply the measure to   
characterize the \textquotedblleft effective freezing\textquotedblright of QD present in the anisotropic 
quantum $XY$ model in a transverse field \cite{xy_group}, 
which appears in the description of certain solid-state realizations \cite{solid_xy}, 
as well as in that of controlled laboratory settings~\cite{num_trap_xy,trap_rev,lab_xy}. 
Moreover, the index is capable of detecting the quantum phase transition in the model. The corresponding 
finite size scaling analysis has also been carried out.

The paper is organized as follows. In Section~\ref{qmeasures}, we briefly discuss the 
measures of quantum correlations used in this study and describe the methodology for investigating their dynamics 
in the presence of local noise. In Section~\ref{twoqubit}, the necessary and sufficient conditions for freezing 
of quantum correlations for bipartite states are derived and a
complementarity of the value of the frozen quantum correlation with the freezing interval is obtained. The 
possible correspondence of the entanglement properties of the initial states to their freezing behavior is 
also addressed. Section~\ref{multiqubit} deals 
with the generalization of the bipartite freezing behavior into multipartite cases. We discuss the phenomenon of 
effective freezing in Section~\ref{efreez}. There, we also quantify the quality of freezing by proposing a freezing index, 
and demonstrate its  behavior in the case of the well-known transverse-field anisotropic $XY$ model. 
Section~\ref{conclude} contains the concluding remarks. 

\section{Quantum correlations: Definitions and dynamics}
\label{qmeasures}

In this section, we provide a brief description of the quantum correlation measures used in the 
paper, namely, the QD and 
the  QWD. We also present the methodology for investigating the dynamics of quantum correlations, 
when quantum states are subjected to local noise, and discuss the freezing of quantum correlations.

\subsection{Quantum discord}
In classical information theory, mutual information between two random variables $A$ and $B$ is defined as 
\begin{eqnarray}
I(A:B) &=& H(A)+H(B)-H(A, B) \nonumber \\
 &=& H(B) - H(B|A),
\end{eqnarray}
where $H(A) = -\sum_{i} p_{i}^{A} \log_{2}p_{i}^{A}$ is the Shannon entropy of $A$, and similarly for $H(B)$,
while $H(A, B)$ is the joint entropy of $A$ and $B$. Here $H(B|A) = H(A, B) - H(A)$ is the conditional entropy of $B$ 
given $A$. Translation of these definitions into the quantum regime of bipartite quantum states leads to two 
inequivalent definitions of mutual information. One of them, which may be identified as the \textquotedblleft total
correlation\textquotedblright\; of a bipartite quantum system $\rho_{AB}$, can be defined as~\cite{tot_corr_group}
\begin{eqnarray}
 I=S\left(\rho_{A}\right)+S\left(\rho_{B}\right)-S\left(\rho_{AB}\right),
 \label{mutual1}
\end{eqnarray}
where $\rho_{A}$ and $\rho_{B}$ are the states of the subsystems $A$ and $B$ respectively and  
$S\left(\rho\right)=-\mbox{Tr}\left[\rho\log_{2}\rho\right]$ is the von Neumann entropy of the quantum state 
$\rho$. An alternative definition of mutual information in the quantum regime takes the form~\cite{disc_group}
\begin{eqnarray}
 J_{\rightarrow}=S\left(\rho_{B}\right)-S\left(\rho_{B}|\rho_{A}\right),
 \label{mutual2}
\end{eqnarray}
where the sign `$\rightarrow$' indicates that the measurement is being performed at $A$. 
Here $S(\rho_{B}|\rho_{A})=\sum_{k}p_{k}S\left(\rho_{AB}^{k}\right)$ is the measured quantum conditional entropy, where  
\begin{eqnarray}
\rho_{AB}^{k}=\left(\Pi_{k}^{A}\otimes I_{B}\right)\rho_{AB}\left(\Pi_{k}^{A}\otimes I_{B}\right)/p_{k}
\label{aftermeasure}
\end{eqnarray}
and 
\begin{eqnarray}
p_{k}=\mbox{Tr}\left[\left(\Pi_{k}^{A}\otimes I_{B}\right)\rho_{AB}\right],
\label{probnorm}
\end{eqnarray}
 with $\{ \Pi_{k}^{A} \}$ being a complete set of rank-$1$ projective measurement and $I_{B}$ denotes the 
identity operator on the Hilbert space of $B$. If $\rho_A$ is a single-qubit state, the rank-$1$ projectors are of the form
$\Pi_k^A=|\Phi_k\rangle\langle\Phi_k|$, $k=1,2$, where
\begin{eqnarray}
|\Phi_1\rangle &=& \cos \frac{\theta}{2}|0\rangle + e^{i \phi} \sin\frac{\theta}{2} |1\rangle\nonumber \\
|\Phi_2\rangle &=& -e^{-i \phi} \sin\frac{\theta}{2} |0\rangle+\cos\frac{\theta}{2}|1\rangle,
\label{pi12}
\end{eqnarray}
with $0\leq\theta\leq\pi$ and $0\leq\phi<2\pi$, and with $\{|0\rangle,|1\rangle\}$ forming the computational 
basis of the qubit Hilbert space.
The classical correlation (CC), $C_{\rightarrow}$, in the state $\rho_{AB}$, between the subsystems $A$ and $B$ 
can be quantified by maximizing $J_{\rightarrow}$ with respect to $\{\Pi_{k}^{A}\}$~\cite{disc_group}, and is given by  
\begin{eqnarray}
 C_{\rightarrow}&=&S\left(\rho_{B}\right)-\underset{\left\{\Pi_{A}^{k}\right\}}{\min}\sum_{k}p_{k}S\left(\rho_{AB}^{k}\right).
 \label{classical}
\end{eqnarray}
The difference between the total correlations and the CC provides the measure of quantum correlation, QD, and 
is given by 
\begin{eqnarray}
 D_{\rightarrow}&=&S\left(\rho_{A}\right)-S\left(\rho_{AB}\right)
 +\underset{\left\{\Pi_{k}^{A}\right\}}{\min}\sum_{k}p_{k}S\left(\rho_{AB}^{k}\right).
 \label{discordexp}
\end{eqnarray}
There is an inherent asymmetry in the definition of the QD which implies that the value of the QD is not 
invariant with respect to the swapping of the parties. Throughout the paper, 
we denote the  CC and the QD by $C$ and $D$ respectively, and calculate QD by performing the measurement on $A$. 
The optimization involved in the definition 
makes the analytical calculation of the QD, for an arbitrary bipartite state, a hard problem. However, for states with 
certain symmetries, such as the Bell-diagonal (BD) states, the optimization is known exactly~\cite{luo_pra_2008}.

\subsection{Quantum work deficit}

The other information-theoretic quantum correlation measure that we consider is the QWD~\cite{wdef_group}. 
It is quantified as the difference between the amount of pure states extractable under suitably restricted 
global and local operations. For a bipartite state $\rho_{AB}$, we consider the class of global operations, termed 
\textquotedblleft closed operations\textquotedblright\; (CO), 
consisting of \textit{(i)} unitary operations, and \textit{(ii)} dephasing the bipartite state by a
set of projectors, $\{\Pi_{k}\}$, defined  on the Hilbert space $\mathcal{H}$ of $\rho_{AB}$. One can show that the amount of pure 
states extractable from $\rho_{AB}$ under CO is given by
\begin{eqnarray}
I_{\mbox{\scriptsize CO\normalsize}}=\log_{2}\mbox{dim}\left(\mathcal{H}\right)-S(\rho_{AB}). 
\label{co}
\end{eqnarray}
On the other hand, the class of  \textquotedblleft closed local operations and 
classical communication\textquotedblright\; (CLOCC) consists of \textit{(i)} local unitary operations,  \textit{(ii)} 
dephasing by local measurement on the subsystem $A$, and \textit{(iii)} communicating the dephased 
subsystem to the other party, $B$, over a noiseless quantum channel. The average quantum state after the projective 
measurement $\{\Pi_{k}^{A}\}$ on $A$ is  $\rho_{AB}^{\prime}=\sum_{k} p_{k}\rho_{AB}^{k}$ 
where $\rho_{AB}^{k}$ and $p_{k}$ are given by Eqs. (\ref{aftermeasure}) and (\ref{probnorm}), respectively.
The amount of pure states extractable under CLOCC is given by 
\begin{eqnarray}
 I_{\mbox{\scriptsize CLOCC\normalsize}}=\log_{2}\mbox{dim}\left(\mathcal{H}\right)-\underset{\left\{\Pi_{k}^{A}\right\}}{\min}
 S\left(\rho_{AB}^{\prime}\right).
 \label{clocc}
\end{eqnarray}
The (one-way) QWD, $W$, is then defined as $W = I_{\mbox{\scriptsize CO\normalsize}}\left(\rho_{AB}\right) 
 - I_{\mbox{\scriptsize CLOCC\normalsize}}\left(\rho_{AB}\right)$.

\subsection{Local dynamics of quantum correlations}

We will consider the situation where each qubit of a multi-qubit system interacts with an independent reservoir 
via decoherence channels. Following the Kraus operator formalism, the density matrix of a system of $N$ 
qubits, $\rho_{0}$, evolves with time as 
\begin{eqnarray}
 \rho=\sum_{k_{1},\cdots,k_{N}}\tau_{k_{1}}^{(1)}\otimes\cdots\otimes\tau_{k_{N}}^{(N)}
 \rho_{0}{\tau_{k_{1}}^{(1)}}^{\dagger}\otimes\cdots\otimes{\tau_{k_{N}}^{(N)}}^{\dagger}, \nonumber \\
 \label{dynamics}
\end{eqnarray}
where $\{\tau_{k_{\alpha}}^{(\alpha)}\}$  describes the noisy channel acting on the qubit 
$\alpha$. The quantum channels describing the interaction of a qubit and its environment (the reservoir) 
can be of various types. We focus on three types of decoherence channels, namely, 
the bit-flip (BF), the phase-flip (PF), and the bit-phase-flip (BPF) channels. The Kraus operators for these channels are given by  
\begin{eqnarray}
 \tau_{0}=\sqrt{1-\frac{\gamma}{2}}I_{2},\;\tau_{1}^{i}=\sqrt{\frac{\gamma}{2}}\sigma^{i},
 \label{kraus}
\end{eqnarray}
where  $\sigma^{i},\,i=1,\,2,\,3$, are the Pauli spin matrices and $I_{2}$ is the identity operator on the 
qubit Hilbert space. Here $\tau_{1}^{i}$, for $i=1,2$, and $3$ correspond to the BF, BPF and PF channels 
respectively. The decoherence probability, $\gamma$, 
called parametrized time, depends explicitly on time $t$ and is taken to be the same for all the qubits with 
$0\leq \gamma\leq 1$. It is convenient to describe the dynamical evolution of systems under decoherence channels 
in terms of the decoherence probabilities,  since such description takes into account a wide range of physical 
situations. A particularly important class of physical scenarios is the one for which 
the functional dependence of the decoherence probability on time, $t$, describes the Markovian approximation, 
where $\gamma$ is often an increasing function, $f_{\Gamma}(t)$, of $t$, such as $\gamma=1-e^{-\Gamma t}$, with  
$\Gamma$ being a \textquotedblleft decay rate\textquotedblright of the function~\cite{yu_prl_2006}. When a non-Markovian 
environment is considered, $\gamma$ may be an oscillatory function of time with a decaying amplitude. One should 
note that under Markovian approximation, $\Gamma\neq 0$ and is fixed for a given bunch of local  environments, 
with the corresponding evolution being scanned by  varying $\gamma$. Once the time evolved density matrix, $\rho^{(\gamma)}$, 
is known, one can calculate its different quantum correlations as functions of system parameters and the decoherence 
probability, to investigate their dynamical behavior under various decoherence channels. 

\subsection{Freezing} 
Even as quantum correlations continue to be regarded as fragile quantities, there are undespairing efforts to control
such decay.
An interesting possibility is the identification of shared quantum states 
that offer a stagnant or near-constant behavior of quantum correlations when affected by noisy time-dynamics. In 
case we find that a quantum correlation measure is a constant in time for a certain time interval, we say that it 
is exhibiting the phenomena of freezing.

\section{Freezing in two-qubit systems}
\label{twoqubit}

A general two-qubit state can be written, up to local unitary 
transformations~\cite{luo_pra_2008,fano_rmp_1983}, as 
\begin{eqnarray}
 \rho_{AB}&=&\frac{1}{4}[I_{A}\otimes I_{B}
 +\sum_{\alpha=1}^3c_{\alpha\alpha}\sigma_{A}^{\alpha}\otimes\sigma_{B}^{\alpha}\nonumber \\
 &&+\sum_{\alpha=1}^3c_{\alpha 0}\sigma_{A}^{\alpha}\otimes I_{B}
 +\sum_{\beta=1}^3c_{0\beta}I_{A}\otimes \sigma_{B}^{\beta}],
 \label{twoqubitequiv}
\end{eqnarray}
where the diagonal correlators, $c_{\alpha \alpha} = \mbox{Tr}[\sigma^{\alpha} \otimes \sigma^{\alpha} \rho_{AB}]$, represent 
\textquotedblleft classical\textquotedblright\; correlators, the single-qubit quantities, 
$c_{\alpha 0} = \mbox{Tr}[\sigma^{\alpha} \otimes I_{B} \rho_{AB}]$ 
and $c_{0 \beta} = \mbox{Tr}[I_{A} \otimes \sigma^{\beta} \rho_{AB}]$, are the magnetizations, and 
$I_{A}$ and $I_{B}$ are identity operators on the Hilbert spaces of $A$ and $B$ respectively.

To address the question of freezing of quantum correlations in bipartite states under local noise, 
we first focus our attention on the BF channel. Using Eq. (\ref{dynamics}), it is easy to show that during 
the BF evolution of $\rho_{AB}$, $c_{11}$, $c_{10}$, and $c_{01}$ remain unchanged,  whereas the correlators 
$c_{\alpha\alpha}$ and the magnetizations 
$c_{0\alpha}$ and $c_{\alpha 0}$ ($\alpha=2,3$) decay with $\gamma$ as $(1-\gamma)^{2}$ and $(1-\gamma)$ respectively. 
To observe freezing phenomena of quantum correlation measures under the BF channel, it is therefore reasonable to choose 
the bipartite state of the form
\begin{eqnarray}
 \rho_{AB}&=&\frac{1}{4}[I_{A}\otimes I_{B}
 +\sum_{\alpha=1}^{3}c_{\alpha\alpha}\sigma_{A}^{\alpha}\otimes\sigma_{B}^{\alpha}\nonumber \\
 && +\left(c_{10}\sigma_{A}^{1}\otimes I_{B}
 +c_{01}I_{A}\otimes\sigma_{B}^{1}\right)],
 \label{twoqubitstate}
\end{eqnarray}
with $c_{\alpha \alpha} \neq 0$ as the initial state of the quantum evolution. We refer to these states as the canonical 
initial states. Next, we will show that to preserve quantum correlation from decohering, inhomogeneous magnetizations play 
an important role. 
The analysis is henceforth mainly carried out for the local BF channel. 
However, a straightforward generalization of the presented results is possible for other local quantum channels, in particular, 
the PF and the BPF channels. 

\subsection{Freezing of QD} 
We begin by investigating the freezing dynamics of quantum correlations, as measured by the QD, 
using the canonical initial states. Unlike the Bell-diagonal (BD) states \cite{luo_pra_2008}, 
QD of the states evolved from CI states cannot be computed analytically \cite{num-err_group}. However, numerical simulations 
show that for a large fraction of states -- \emph{special} CI (SCI) states $(\mathcal{S}_{1})$ --
the optimization takes place for the projectors 
corresponding to three sets of \textquotedblleft regular\textquotedblright\; values 
$\{\theta,\phi\}$:  $s_{1} = \{\theta=0,\pi\}$, $s_{2} = \{\theta=\pi/2,\phi=\pi/2,3\pi/2\}$, and 
$s_{3}=\{\theta=\pi/2,\phi=0,\pi\}$. 
The existence of the complementary class, which we denote by $\mathcal{S}_{2}$, makes the analytical calculation 
of the QD for $\rho_{AB}$ difficult. If $D$ denotes the QD of the state $\rho_{AB}$ and $D'$ represents the QD 
calculated with the assumption that $\rho_{AB} \in \mathcal{S}_{1}$, then our numerical analysis shows that $\epsilon<0.0028$,  
where $\epsilon=\max\{D^{\prime}-D\}$ is the maximum value of the error due to the assumption. 
Similar findings have been reported earlier
for two-qubit $X$ states \cite{num-err_group}. 
For the numerical simulation, the two-qubit canonical initial state, 
$\rho_{AB}$, is generated  on a grid with a separation of $\sim10^{-3}$ for all correlators, $c_{\alpha \alpha}$, 
and magnetizations, $c_{10}$ and $c_{01}$.  
Proposition I provide a necessary and sufficient criterion for freezing of QD for the SCI  
states. Numerical evidence strongly suggests that the proposition holds for the entire 
class of CI states up to the second decimal place.

\noindent\textbf{Proposition I.} \textit{
If a two-qubit SCI state is sent through local BF channels, an NS condition for the QD in 
the evolved state to remain constant over a finite interval of time is given by either of the following sets 
of equations:}
\begin{eqnarray}
   \begin{cases}
     (i)  & (c_{22}/c_{33}) = -(c_{10}/c_{01}) = -c_{11},\\
     (ii) & c_{33}^2 + c_{01}^2  \leq  1,\\
     (iii)& F\left( \sqrt{c_{33}^2 + c_{01}^2}\right) \leq  F(c_{11}) + F (c_{01}) - F(c_{10});
   \end{cases}\nonumber\\   
\label{necsuf1}
\end{eqnarray}
\begin{eqnarray}
   \begin{cases}
     (i)  & (c_{33}/c_{22}) = -(c_{10}/c_{01}) = -c_{11},\\
     (ii) & c_{22}^2 + c_{01}^2  \leq  1,\\
     (iii)& F\left( \sqrt{c_{22}^2 + c_{01}^2}\right) \leq  F(c_{11}) + F (c_{01}) - F(c_{10}).
   \end{cases}\nonumber \\
\label{necsuf2}
\end{eqnarray}
\textit{Here, $F(y)=2\left(H(\frac{1+y}{2})-1\right)$, with $H(\alpha) = -\alpha \log_{2}\alpha - (1-\alpha)\log_{2}(1-\alpha)$ 
being the binary entropy function.}

\noindent Note: We call the function $F$ as the \textquotedblleft freezing entropy\textquotedblright\; 
and the relations (\ref{necsuf1})(iii) and (\ref{necsuf2})(iii) as the 
\textquotedblleft freezing subadditivity\textquotedblright\; I and II, for the QD, respectively.

\noindent\textbf{Proof.} 
For a state $\rho_{AB}\in\mathcal{S}_{1}$, QD is given by $D=\min\{D_{l}\}$ where $l=1,2,$ and $3$ 
correspond to the sets $s_1$, $s_2$, and $s_3$, respectively, with 
\small 
\begin{eqnarray}
 D_{l}&=&S(\rho_{A})-S(\rho_{AB}^{(\gamma)})-\sum_{i}p_{i}\sum_{ij}\chi_{ij}\log_{2}\chi_{ij}\delta_{l3}\nonumber \\
 &&+(1-\delta_{l3})(1+F(c^{\prime}_{\delta})/2).
 \label{dbig}
\end{eqnarray}\normalsize
Here, $\delta_{ll{'}}$ denote the Kronecker delta, ${c^{\prime}_{\delta}}^{2}=c_{01}^{2}+(1-\gamma)^{4}(c_{33}^{2}\delta_{l1}+
 c_{22}^{2}\delta_{l2})$, $p_{i}=\frac{1}{2}\left(1+(-1)^{i}c_{10}\right)$, and 
$ \chi_{ij}=
 (1+(-1)^{i}c_{10}+(-1)^{j}(c_{01}+(-1)^{i}c_{11}))
/2\left(1+(-1)^{i}c_{10}\right)$.
Note that the marginal states 
$\rho_{A}$ and $\rho_{B}$ of the canonical initial state $\rho_{AB}$ do not vary with $\gamma$. 
Let us first focus on the necessity of the conditions given in (\ref{necsuf1}) and 
(\ref{necsuf2}). If freezing of QD takes place, $D$ must be invariant with $\gamma$ for a finite interval. Let us assume that 
$D=D_{1}$, in that interval. From the expression of $D_{1}$, it is easy to show that for $D_{1}$ to be independent of $\gamma$, 
condition (\ref{necsuf1})(i) must be satisfied. 
Under this condition, (\ref{necsuf1})(ii) is required to ensure the positivity of 
the initial state $\rho_{AB}\in\mathcal{S}_{1}$. Since $D_{1}=\min\{D_{l}\}$, $l=1,2,3$, $D_{3}$ must be 
greater that $D_{1}$ which leads to the condition (\ref{necsuf1})(iii), thereby proving the necessity of the group 
of conditions given in (\ref{necsuf1}) for the occurrence of freezing of QD. Next, we assume that $D=D_{2}$. In a 
similar fashion as in the previous case, one can show that the set of conditions given in (\ref{necsuf2}) is 
necessary for the freezing of QD. Lastly, let $D=D_{3}$. From Eq. (\ref{dbig}), it is easy to see that the only term 
dependent on $\gamma$ is $S(\rho_{AB}^{(\gamma)})$. The eigenvalues of the time evolved state $\rho_{AB}^{(\gamma)}$ with 
the state $\rho_{AB}\in\mathcal{S}_{1}$ as the initial state can be easily determined to be 
\begin{eqnarray}
\lambda_{1}&=&\frac{1}{4}(1+c_{11}-\sqrt{(c_{10}+c_{01})^{2}+(\gamma-1)^{4}(c_{22}-c_{33})^{2}}),\nonumber\\
\lambda_{2}&=&\frac{1}{4}(1+c_{11}+\sqrt{(c_{10}+c_{01})^{2}+(\gamma-1)^{4}(c_{22}-c_{33})^{2}}),\nonumber\\
\lambda_{3}&=&\frac{1}{4}(1-c_{11}-\sqrt{(c_{10}-c_{01})^{2}+(\gamma-1)^{4}(c_{22}+c_{33})^{2}}),\nonumber\\
\lambda_{4}&=&\frac{1}{4}(1-c_{11}+\sqrt{(c_{10}-c_{01})^{2}+(\gamma-1)^{4}(c_{22}+c_{33})^{2}}).\nonumber\\
 \label{eigengamma}
\end{eqnarray}
One can easily show that $S(\rho_{AB}^{(\gamma)})$ varies with $\gamma$ for all possible non-zero values of the correlators 
and the magnetizations which is not possible if QD freezes. Hence (\ref{necsuf1}) and (\ref{necsuf2}) are the necessary 
conditions for the occurrence of freezing in QD in the case of initial two-qubit states $\rho_{AB}\in\mathcal{S}_{1}$.

To prove the sufficiency of the conditions, we first consider the set of conditions in (\ref{necsuf1}). If 
condition (\ref{necsuf1})(i) is imposed over the initial two-qubit state $\rho_{AB}\in\mathcal{S}_{1}$, it can be shown that
$D_1$ is independent of time for all values of $\gamma$. 
One should note that a condition similar to this one has earlier been reported for the PF channel~\cite{li_pra_2011}.
For $\left|c_{11}\right|=1$, the initial state is a pure state with QD monotonically decaying with $\gamma$. 
Moreover, $D_{2} > D_{1}$ $\forall$ $\gamma$, when Eq. (\ref{necsuf1})(i) is satisfied implying that the QD 
is given by $D = \min\{D_l\}$ with $l=1,3$. 
Besides (\ref{necsuf1})(i), condition (\ref{necsuf1})(ii) ensures positivity of the initial two-qubit 
state $\rho_{AB}\in\mathcal{S}_{1}$. 
Application of conditions (\ref{necsuf1})(i)-(ii) leads to the following forms of the functions $D_{1}$ and $D_{3}$: 
\begin{eqnarray}
 D_{1}&=&\frac{1}{2}(F(c_{10})-F(c_{11})), \nonumber \\ 
 D_{3}&=&\frac{1}{2}(F(c_{01})-F(c^{\prime})).
 \label{dis_exp}
\end{eqnarray}
Here, ${c^{\prime}}^{2}=c_{01}^2 + c_{33}^2(1-\gamma)^{4}$. Note that $D_{3}$ is a monotonically decreasing function of $\gamma$.  When condition (\ref{necsuf1})(iii) is applied, 
we get $D_{3}>D_{1}$ for a finite interval of time in which QD freezes.
Similarly, one can prove that the QD, given by $D_{2}$, is invariant with $\gamma$ when the sets of conditions 
given in (\ref{necsuf2}) are obeyed. 
Hence for the two-qubit states $\rho_{AB}\in\mathcal{S}_{1}$, the set of conditions (\ref{necsuf1}) and (\ref{necsuf2}) are both necessary and 
sufficient for the QD to remain constant under the BF noise. \hfill $\blacksquare$\\

\begin{figure}
 \includegraphics[scale=0.345]{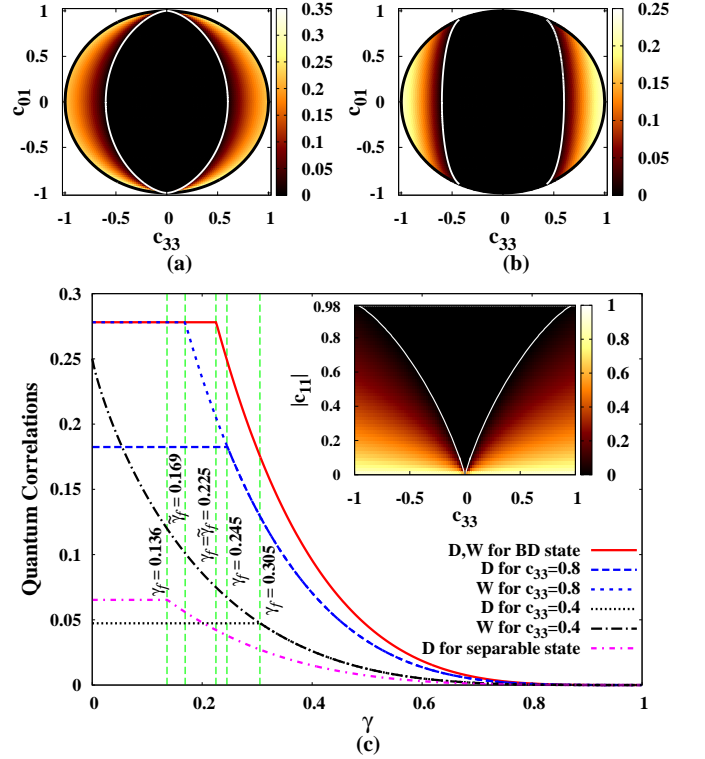}
 \caption{(Color online) 
 Freezing phase diagram. The freezing of (a) QD and (b) QWD under local BF channels for the CI states with $|c_{11}|=0.6$
 and obeying condition (a) (\ref{necsuf1})(i) for QD and (b) (\ref{wnecsuf1})(i) for QWD. See text for details.
 (c) The dynamics of quantum correlations, as measured by the QD and the QWD, using two-qubit CI states obeying conditions 
 (\ref{necsuf1}) or (\ref{wnecsuf1}) with $c_{33}^2 + c_{01}^2 = 1$. For all of these states, the value of  $|c_{11}|=0.6$. 
 Inset: Another freezing phase diagram for QWD on the $(c_{33},|c_{11}|)$ plane for the CI states obeying  
 condition (\ref{wnecsuf1})(i)-(iii) with $c_{33}^2 + c_{01}^2 = 1$.  
 All quantities plotted are dimensionless, except QD, which is in bits, and QWD, which is in qubits.
 }
 \label{twoqubitphase}
\end{figure}

The freezing phase diagram on the $(c_{33},c_{01})$ plane,  for SCI states, is exhibited in 
Fig.~\ref{twoqubitphase}(a) for a fixed value of $|c_{11}|=0.6$.
For given values of $c_{11}$, one obtains different freezing phase diagrams depending on whether condition (\ref{necsuf1}) or (\ref{necsuf2}) is used. Here we chose condition (\ref{necsuf1}) for 
Fig. \ref{twoqubitphase}(a). Then, the states that show freezing of QD under the BF channel are enclosed by the circle 
$c_{01}^{2}+c_{33}^{2}=1$ and also satisfy the freezing subadditivity I for QD. The white region outside the circle 
depicts states that violate positivity.
Freezing occurs, for a finite parametrized time interval , $0\leq\gamma\leq\gamma_{f}$, within the two crescents -- they 
form the \textquotedblleft freezing crescents\textquotedblright\; 
for QD for the chosen parameter space. 
We refer to $\gamma_{f}$ as the ``freezing terminal". The freezing crescents as well as the freezing terminals are functions 
of the input quantum state, the channel, and the measure employed to quantify quantum correlations. 
$\gamma_f$ can be found by solving
\small
\begin{eqnarray}
F(\sqrt{c_{01}^{2} + c_{33}^{2}(1-\gamma)^{4}})
=  F(c_{11}) + F (c_{01}) - F(c_{10}).
\label{gammaf}
\end{eqnarray}
\normalsize
In Fig. \ref{twoqubitphase}(a), the $\gamma_{f}$ are 
mapped onto  the freezing crescents  in the phase diagram. 
The states for which freezing takes place are indicated by the faded regions while the black region represents
states for which the QD decays with $\gamma$. The different shades in 
the freezing crescents indicate the values of the freezing terminal, $\gamma_{f}$.
Note that the states inside the freezing crescents 
can be generated by BF evolution from the states lying on the perimeter of 
$c_{33}^{2}+c_{01}^{2}=1$. If $|c_{11}|$ is decreased, the freezing region expands, thereby 
indicating an increase in $\gamma_{f}$ for fixed $c_{33}$ and $c_{01}$, although the value 
of the frozen QD decreases. We revisit this issue in Proposition IV.
Note that choosing condition (\ref{necsuf2}) to draw the freezing phase diagram, the corresponding 
$\gamma_{f}$ would be given by the equation obtained by replacing $c_{33}$ by $c_{22}$ Eq. (\ref{gammaf}).

Let us now state two corollaries which follow directly from Proposition I.

\noindent\textit{\textbf{Corollary 1.}} \textit{When an SCI state,
satisfying the NS freezing conditions for QD, is subjected to 
local BF noise, the freezing terminal attains its maximum value for given values of $c_{11}$ and $c_{01}$, 
at the maximum allowed value of $\left|c_{33}\right|$ or $\left|c_{22}\right|$.} 

\noindent\textbf{Proof.} From Eqs. (\ref{necsuf1}) and (\ref{gammaf}), 
$c_{01}^2 + c_{33}^2 (1-\gamma_f)^4 = \textrm{constant}$ for fixed $c_{11}$ and $c_{01}$. 
This implies that $\gamma_f$  attains its maximum value for $|c_{33}|_{max} =\sqrt{1 - c_{01}^2}$. A similar proof exists if Eq. (\ref{necsuf2}) is considered instead of Eq. (\ref{necsuf1}). 
\hfill $\blacksquare$\\

\noindent\textit{\textbf{Corollary 2.}} \textit{When an SCI state
is subjected to local BF noise, the QD will always decay if the magnetization is 
homogeneous.} 

\noindent\textbf{Proof.} Homogeneity of magnetization implies $c_{01}=c_{10}$, and from Eq. (\ref{necsuf1})(i) or (\ref{necsuf2})(i), it is clear that
the homogeneity of non-zero magnetization requires $\left| c_{11}\right|=1$  which violates the 
necessary condition for freezing of QD in SCI states. 
\hfill $\blacksquare$\\

Note that if $c_{01}=c_{10}=0$, the CI state reduces to a BD state, in which freezing of 
different quantum correlations occur
~\cite{mazzola_prl_2010,adesso_pra_2013,disc_freez_group,disc_expt_group}.
In the case of the BD states, the second relation in (\ref{necsuf1})(i) (or in (\ref{necsuf2})(i)) does not hold while the first condition is 
still valid and gives a necessary condition for freezing~\cite{mazzola_prl_2010,adesso_pra_2013}.
In Fig. \ref{twoqubitphase}(a), for $c_{01}=0$, the BD states are along the horizontal diameter of the 
circle. The two end points of that diameter represent BD states for which QD is known to exhibit freezing~\cite{mazzola_prl_2010}. 
The dynamics of QD for the BD state 
with $|c_{11}|=|c_{22}|=0.6$, $|c_{33}|=1$, $c_{01}=c_{10}=0$ is shown in Fig. \ref{twoqubitphase}(c). 
Note that there exist initial states, eg. the states satisfying (\ref{necsuf1}) and lying on 
$c_{33}^{2}+c_{01}^{2}=1$, for which freezing terminals longer than that of 
the BD state can be achieved (Fig. \ref{twoqubitphase}(c)). Identifying such a state with a prolonged 
constancy of QD under decoherence can be 
of vital importance in realizing quantum information protocols.

The results mentioned above are only for the SCI states that 
satisfy conditions (\ref{necsuf1})(i)-(iii). Our numerical findings suggest that 
a small fraction of the states that obey conditions (\ref{necsuf1})(i) and (ii)
belong to the set $\mathcal{S}_{2}$. 
Extensive numerical simulations indicate that such states are found only in 
the regions on the freezing phase diagram where the quantum states do not show freezing 
behavior. Irrespective of the optimal sets in the  measurement of QD, 
Proposition I and numerical simulations strongly suggest that the QD of the entire class of CI states, 
would exhibit freezing if and only if they satisfy the conditions (\ref{necsuf1})(i) and (\ref{necsuf1})(ii). 

\subsection{Freezing of QWD} 
We now move on to investigate the freezing phenomena for other information-theoretic quantum correlation measures. 
Freezing of QD has been extensively studied for BD states, for which QWD and QD 
coincide~\cite{wdef_group}. 
Let us consider an arbitrary bipartite state, $\varrho_{AB}$, of which 
$\varrho_{A(B)}$ is the marginal state of the subsystem $A(B)$ obtained by tracing out the other subsystem $B(A)$,  
and advance to the following proposition.
 
\noindent\textbf{Proposition II.} \textit{For a given bipartite state $\varrho_{AB}$ evolving under local BF
channels, if the optimizations in QD and QWD occur in the same optimal ensemble 
$\{\bar{p}_{k},\bar{\varrho}_{AB}^{k}\}$ for $\gamma\leq\gamma_{f}$, and if 
$H\left(\left\{\bar{p}_{k}\right\}\right)-S\left(\varrho_{A}\right)$ is independent of time for the same 
interval, then QWD freezes for $\gamma\leq\gamma_{f}$ provided QD freezes for 
$\gamma\leq\gamma_{f}^{0}$ where $\gamma_{f}^{0}\geq\gamma_{f}$.}

\noindent\textbf{Proof.} From the definitions of QD and QWD, we get
\begin{eqnarray}
W&=&D-S\left(\varrho_{A}\right)+
\underset{\left\{\Pi_{A}^{k}\right\}}{\mbox{min}}\sum_{k}p_{k}S\left(\varrho_{AB}^{k}\right)\nonumber \\
&&-\underset{\left\{\Pi_{A}^{k}\right\}}{\mbox{min}}S\left(\sum_{k}p_{k}\varrho_{AB}^{k}\right).  
\label{discorddeficit}
\end{eqnarray}
Using the concavity of von Neumann entropy,
\begin{eqnarray}
 \sum_{k}\bar{p}_{k}S\left(\bar{\varrho}_{AB}^{k}\right)
 =S\left(\sum_{k}\bar{p}_{k}\bar{\varrho}_{AB}^{k}\right)+H\left(\left\{\bar{p}_{k}\right\}\right),
\end{eqnarray}
using which, we reach
\begin{eqnarray}
 W=D-S\left(\varrho_{A}\right)+H\left(\left\{\bar{p}_{k}\right\}\right),
 \label{frozendeficit}
\end{eqnarray}
provided both the minimizations in Eq. (\ref{discorddeficit}) take place for the same ensemble. 
If Eq. (\ref{frozendeficit}) is satisfied, the freezing of QWD demands the 
freezing of  QD, provided  $H\left(\left\{\bar{p}_{k}\right\}\right)-S\left(\rho_{A}\right)$ is constant in 
time in the relevant interval. Note that the result is not restricted to two-qubit states. \hfill $\blacksquare$

For the SCI states, the conditions in the above proposition can be relaxed. 
Specifically, we obtain the following corollary.

\noindent\textit{\textbf{Corollary 3.}} \textit{When an SCI state is sent through 
local BF channels, QWD freezes whenever QD shows 
freezing behavior, provided the optimizations occur for the same ensemble.}

\noindent\textbf{Proof.} For an SCI state, $\rho_{AB}$,  $S(\rho_{A})$ 
remains unaltered with time. From the relation between QD and QWD given in Eq. (\ref{frozendeficit}), for  
$\rho_{AB}$, we find that $\bar{p}_{k} = 1/2$ $\forall k$ whenever QD freezes and hence the proof.
\hfill $\blacksquare$\\ 

Similarly as for QD, numerical investigation shows that also in the case of QWD, there exist two sets 
of states, $\tilde{\mathcal{S}}_{1}$ and $\tilde{\mathcal{S}}_{2}$, depending on the optimal measurements.
For the states $\rho_{AB}\in\tilde{\mathcal{S}}_{1}$, the optimization of QWD takes place for the projectors 
corresponding to three sets of ``regular'' values, while the rest of the states constitute the set $\tilde{\mathcal{S}}_{2}$.  
Interestingly, for the states $\rho_{AB}\in\tilde{\mathcal{S}}_{1}$, the three regular sets  
are identical to those for QD. We also observe that the set of states, for which 
the optimal measurements are at irregular values, is small. Let us now state 
the NS condition for the freezing behavior of QWD.

\noindent\textbf{Proposition III.} \textit{
If a two-qubit state in $\tilde{\mathcal{S}}_{1}$ is sent 
through local BF channels, an NS condition for QWD in the evolved state to remain 
constant over a finite interval of time is given by either of the following sets of equations:}
\begin{eqnarray}
   \begin{cases}
     (i)  & \frac{c_{22}}{c_{33}} = -\frac{c_{10}}{c_{01}} = -c_{11},\\
     (ii) & c_{33}^2 + c_{01}^2  \leq  1,\\
     (iii)& F\left( \sqrt{c_{33}^2 + c_{01}^2}\right) \leq  F(c_{11}) + F (c_{01}));
   \end{cases}\nonumber \\
\label{wnecsuf1}
\end{eqnarray}
\begin{eqnarray}
   \begin{cases}
     (i)  & \frac{c_{33}}{c_{22}} = -\frac{c_{10}}{c_{01}} = -c_{11},\\
     (ii) & c_{22}^2 + c_{01}^2  \leq  1,\\
     (iii)& F\left( \sqrt{c_{22}^2 + c_{01}^2}\right) \leq  F(c_{11}) + F (c_{01})).
   \end{cases}\nonumber \\
\label{wnecsuf2}
\end{eqnarray}

\noindent\textbf{Proof.} Proceeding in a similar fashion as in the case of QD, it can be shown that QWD 
of the time evolved two-qubit state, $\rho_{AB}^{(\gamma)}$, is given by $W=\min\{W_{l}\}$ with $l=1,2,$ and $3$ corresponding to the three 
sets of $\{\theta,\phi\}$ values, $s_{1}$, $s_{2},$ and $s_{3}$, where
\begin{eqnarray}
 W_{l}&=&2(\delta_{l1}+\delta_{l2})-S(\rho_{AB}^{(\gamma)})-\delta_{l3}\sum_{i=1}^{4}\lambda_{i}\log_{2}\lambda_{i}\nonumber \\
 &&+\frac{1}{2}F\left(\sqrt{c_{01}^{2}+(c_{33}^{2}\delta_{l1}+c_{22}^{2}\delta_{l2})(1-\gamma)^{4}}\right).
\end{eqnarray}
Here, 
\begin{eqnarray}
 \lambda_{1}&=&\frac{1}{4}\left(1+c_{01}+c_{10}+c_{11}\right),\nonumber \\
 \lambda_{2}&=&\frac{1}{4}\left(1-c_{01}-c_{10}+c_{11}\right),\nonumber \\
 \lambda_{3}&=&\frac{1}{4}\left(1-c_{01}+c_{10}-c_{11}\right),\nonumber \\
 \lambda_{4}&=&\frac{1}{4}\left(1+c_{01}-c_{10}-c_{11}\right).\nonumber \\
\end{eqnarray}

 We begin with the proof for the necessity of the conditions (\ref{wnecsuf1}) and \ref{wnecsuf2}. First, let us assume that 
the QWD is given by $W_{1}$. For freezing to occur, $W_{1}$ must be independent of $\gamma$ in a finite interval. It is easy to 
show that if $W_{1}$ is independent of $\gamma$, then condition (\ref{wnecsuf1})(i) is satisfied. Then to ensure the positivity 
of the initial state, condition (\ref{wnecsuf1})(ii) must be satisfied. Also, $W=W_{1}$ implies that $W_{3}>W_{1}$ for a 
finite range of $\gamma$ leading to the condition (\ref{wnecsuf1})(iii). Hence the set of conditions (\ref{wnecsuf1}) is 
necessary for freezing of QWD when $W=W_{1}$. In a similar fashion, one can show that the set of conditions (\ref{wnecsuf2})(i)-(iii) is necessary for freezing of QWD when $W=W_{2}$. For $W=W_{3}$, similar to the case of the QD, the 
$\gamma$-dependence comes through the term $S(\rho_{AB}^{(\gamma)})$ which can be determined using the eigenvalues given 
in Eq. (\ref{eigengamma}). One can easily show that the function $W_{3}$ always depends on $\gamma$ for all possible 
values of the correlators and magnetizations,
thereby proving that freezing of QWD is not possible for $W=W_{3}$. Hence the sets of conditions given in  (\ref{wnecsuf1}) 
and (\ref{wnecsuf2}) are necessary for freezing to occur in the case of QWD with $\rho_{AB}\in\tilde{\mathcal{S}}_{1}$
as initial states. 

To prove the sufficiency of the conditions, we start with the set of conditions (\ref{wnecsuf1}). When condition (\ref{wnecsuf1})(i)
is imposed, the function $W_{1}$ is independent of $\gamma$, and $W_{2}>W_{1}$, 
implying $W=\min\{W_{l}\}$ with $l=1,3$. The second condition of (\ref{wnecsuf1}) is required to ensure positivity of the 
initial state once the condition (\ref{wnecsuf1})(i) is applied.  
Under the conditions (\ref{wnecsuf1})(i) and (ii), $W_{1}=-\frac{1}{2}F(c_{11})$ whereas 
$W_{3}=\frac{1}{2}(F(c_{01})-F(\sqrt{c_{01}^{2}+c_{33}^{2}(1-\gamma)^{4}}))$, which 
decreases monotonically with $\gamma$. If the third condition of  
(\ref{wnecsuf1}) is applied, $W_{3}>W_{1}$ for a finite range of $\gamma$ so that $W=W_{1}$ in that range. 
Since $W_{1}$ is invariant with $\gamma$, freezing of QWD takes place in that range thereby proving the sufficiency of the 
set of conditions (\ref{wnecsuf1}). Following a similar path, one can show that $W=W_{2}$ freezes for a finite interval of 
$\gamma$, when conditions (\ref{wnecsuf2}) are applied. \hfill $\blacksquare$\\

\noindent \textbf{Comparison.} There are clear signatures that point to differences in the behavior of QD and QWD in the dynamics. 
In particular, extensive numerical searches show that no CI state  
satisfying (\ref{wnecsuf1})(i) and (ii)  is in $\tilde{\mathcal{S}}_{2}$. This
is in stark contrast to the findings for QD. 
 Like QD, the freezing terminal, $\tilde{\gamma}_{f}$, for QWD can be determined as the solution of the equation $W_{1}=W_{3}$ 
(assuming conditions (\ref{wnecsuf1})). The freezing of the QWD is depicted in the $(c_{33},c_{01})$ plane in 
Fig. \ref{twoqubitphase}(b) for CI states with $|c_{11}|=0.6$ and 
when the conditions (\ref{wnecsuf1})(i) and (ii) are obeyed. For fixed parameters, the freezing 
region for QWD can be smaller than that of QD, indicating the existence $\rho_{AB}\in\mathcal{S}_{1}$ for which 
QD freezes but QWD does not. Interestingly, for such states, we find that the optimal projectors 
are different for QD and  QWD. Eg., see Fig. \ref{twoqubitphase}(c) for $c_{33}=0.4$. The inset of 
Fig. \ref{twoqubitphase}(c) maps the $\gamma_{f}$ for the QWD in the $(|c_{11}|,c_{33})$ plane under 
condition (\ref{wnecsuf1}) with $c_{01}^2 + c_{33}^2 =1$. 
The shades represent similar situations as in the case 
of Fig. \ref{twoqubitphase}(a). The black inner region between the two curves correspond to states for which the QWD decay 
monotonically under the BF noise and exhibit no freezing.  
Contrary to the behaviour of QWD, QD shows freezing 
for all states on the $(|c_{11}|,c_{33})$ plane under the same condition except at $c_{33}=0$, for which the initial state is 
completely classical. This is an example where the behavior of QD and QWD differ in a very drastic way.
In contrast to earlier findings, focussing on BD states~\cite{adesso_pra_2013}, our analysis 
clearly shows that freezing of quantum correlations depends explicitly on the choice of the correlation measures.

\subsection{Complementarity}

\begin{figure}
 \includegraphics[scale=0.35]{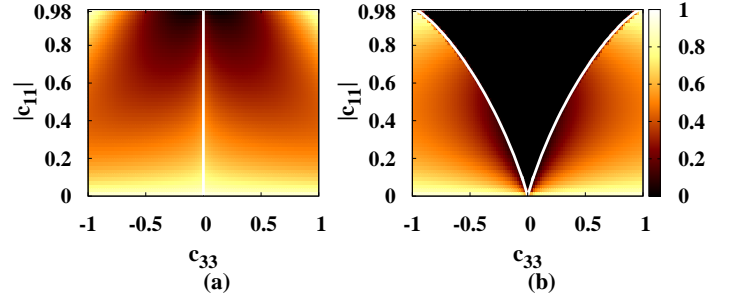}
 \caption{(Color online) Complementarity. Value of $Q_{f}+\gamma_{F}$ for QD (a) and QWD (b) using the 
 canonical initial state  
 obeying Eq. (\ref{necsuf1}) and (\ref{wnecsuf1}) respectively with $c_{33}^2 + c_{01}^2 =1$ as the initial 
 state under BF dynamics. The sum $Q_{f}+\gamma_{F}$ is represented by different shades, as indicated by the 
 color-bar, on the $(c_{33},|c_{11}|)$ plane. The QD is zero for $c_{33}=0$ (the white 
 vertical line in (a)). For  QWD, $\tilde{\gamma}_{f}=0$ above the white curves in (b), as shown earlier (inset of Fig. 
 \ref{twoqubitphase}(c)). The dimensions are the same as in Fig. \ref{twoqubitphase}.}
 \label{comple}
\end{figure}

From the freezing behavior of QD and QWD, we observe that for the CI states, the frozen 
values of the quantum correlation measures increase while the freezing terminals decrease with the tuning of appropriate system parameters. 
This observation is made more precise in Proposition IV. 

\noindent\textbf{Proposition IV.} \textit{If a two-qubit BD state freezes under  
local BF noise, the frozen quantum correlation $Q_{f}$, as
measured by QD or QWD, and the freezing terminal, $\gamma_{F}$, satisfy the complementarity 
relation}
\begin{eqnarray}
 Q_{f}+\gamma_{F}\leq1,
 \label{comple_rel}
\end{eqnarray}
\textit{where $\gamma_{F}=\gamma_{f}$ or $\tilde{\gamma}_{f}$, respectively.}

\noindent\textbf{Proof.} In the case of the BD states, $c_{10}=c_{01}=0$ in Eq. (\ref{twoqubitstate}), and the 
QD, under condition (\ref{necsuf1}), is given by $D=-\frac{1}{2}F(c_{11})$, so that $\gamma_{f}=1-\sqrt{|c_{11}|/|c_{33}|}$. 
Now, $D+\gamma_{f}$ is an even function of $c_{11}$, having no maxima and a 
single minima between $0$ and $1$. As a function of $c_{33}$, $D+\gamma_{f}$ attains its maximum at $|c_{33}|=1$.
The maximal value of the function is $1$ for $c_{11}=0,\pm 1$. 
Since $W=D$ in the case of the BD state, the proposition holds for QWD as 
well. \hfill $\blacksquare$ \\

Now the question remains whether the complementary relation holds for other classes of states. 
For the CI states that obey condition (\ref{necsuf1}), 
the values $\gamma_{f}$ are obtained by the implicit equation (\ref{gammaf}). 
Similar equations can be solved for the other cases. Numerical analysis with such equations reveal that 
the complementarity relation (\ref{comple_rel}) is valid for all 
possible states satisfying the NS conditions in Proposition I and III, and therefore correspond to 
both QD and QWD. Specifically, we find that the maximum 
of  $Q_{f}+\gamma_{F}$ is $1$, and occurs only when  $c_{11}=0$ or $|c_{11}|=|c_{33}|=1$ (see Fig. \ref{comple}).  

\subsection{Non-convexity}
Up to now, we have concentrated on the conditions on the parameters of the class of states for which QD and QWD freeze. 
We now study the properties of the set of states which show freezing for QD as well 
as those for QWD. In particular, we have the following proposition. 

\noindent\textbf{Proposition V.} \textit{The SCI states 
that exhibit freezing of QD form a non-convex set. The same is true for QWD.}

\noindent\textbf{Proof.} If the sets are convex, then the state 
$\rho=p\rho^{1}_{AB}+(1-p)\rho^{2}_{AB}$ for all $0\leq p\leq 1$ will be a state that will exhibit freezing, if 
$\rho^{1}_{AB}$ and $\rho^{2}_{AB}$ does so. We note that the necessary 
conditions for freezing for both the QD and the QWD are given in (\ref{necsuf1})(i) and (\ref{necsuf2})(ii). Therefore, 
for convexity to hold, we must have the relation  
\begin{eqnarray}
 \frac{pc_{22}^{1}+(1-p)c_{22}^{2}}{pc_{33}^{1}+(1-p)c_{33}^{2}}&=&
 -\frac{pc_{10}^{1}+(1-p)c_{10}^{2}}{pc_{01}^{1}+(1-p)c_{01}^{2}}\nonumber \\
 &=&-pc_{11}^{1}-(1-p)c_{11}^{2},
 \label{nonconvex}
\end{eqnarray}
or the relation 
\begin{eqnarray}
 \frac{pc_{33}^{1}+(1-p)c_{33}^{2}}{pc_{22}^{1}+(1-p)c_{22}^{2}}&=&
 -\frac{pc_{10}^{1}+(1-p)c_{10}^{2}}{pc_{01}^{1}+(1-p)c_{01}^{2}}\nonumber \\
 &=&-pc_{11}^{1}-(1-p)c_{11}^{2}.
 \label{nonconvex2}
\end{eqnarray}
true for all $p$. 
Here, $c_{\alpha \alpha}^1$ and $c_{\alpha \alpha}^2$, $\alpha=0,1,2,3$ denote the correlators 
and magnetizations of $\rho_{AB}^1$ and $\rho_{AB}^2$ 
respectively. For arbitrary values of the correlators, the above equations are not satisfied except for $p=0,1,$ 
proving the non-convexity of the sets. \hfill $\blacksquare$

\begin{figure}
 \includegraphics[scale=0.35]{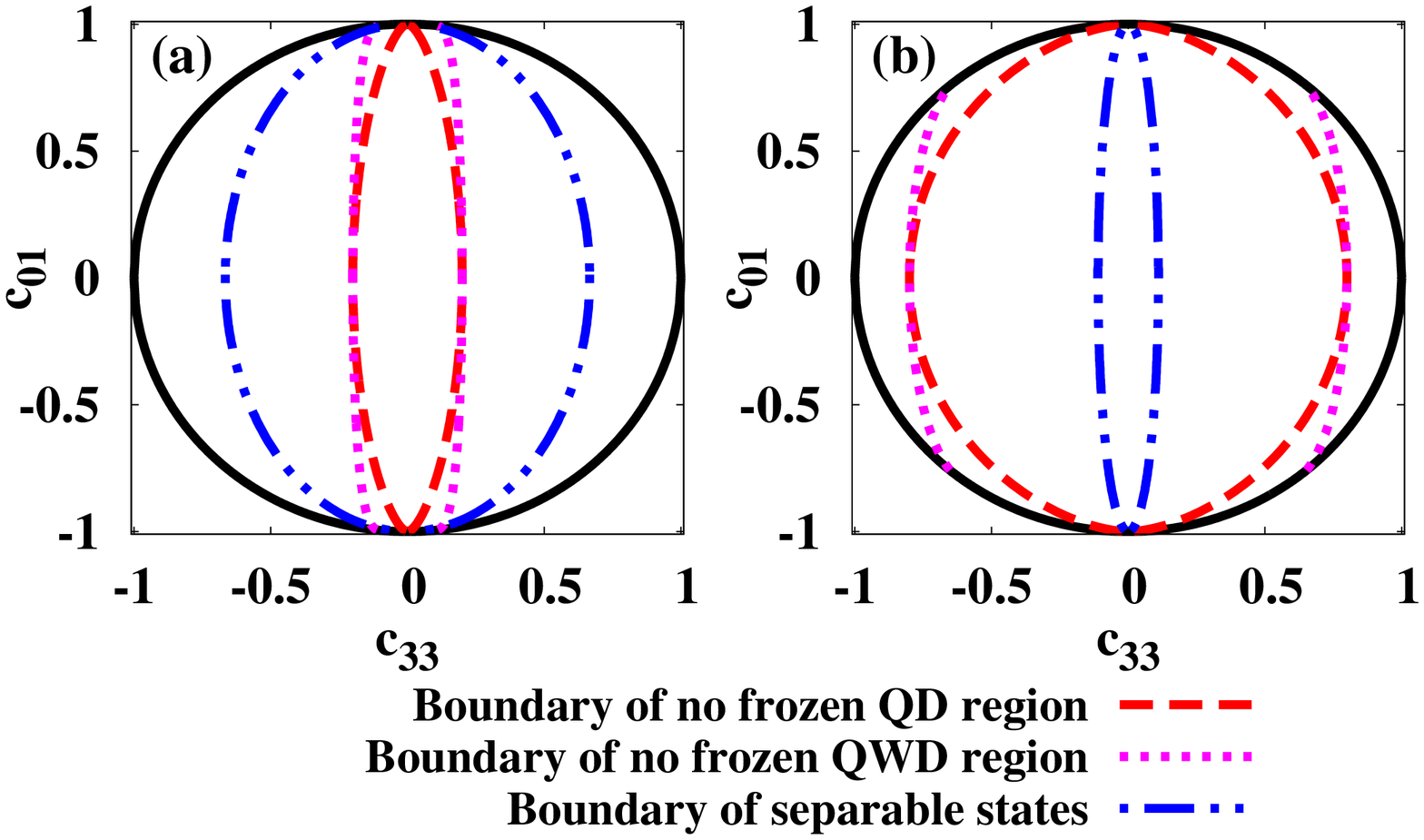}
 \caption{(Color online) The boundaries of the entangled and freezing regions for QD and QWD are plotted 
 for $|c_{11}|=0.2$ (a) and $|c_{11}|=0.8$ (b). Clearly, the entangled region increases as 
 the value of $|c_{11}|$ increases, while the trend is opposite for the freezing regions of QD and 
 QWD. All quantities plotted are dimensionless.}
 \label{entphase}
\end{figure}

\begin{figure}
 \includegraphics[scale=0.4]{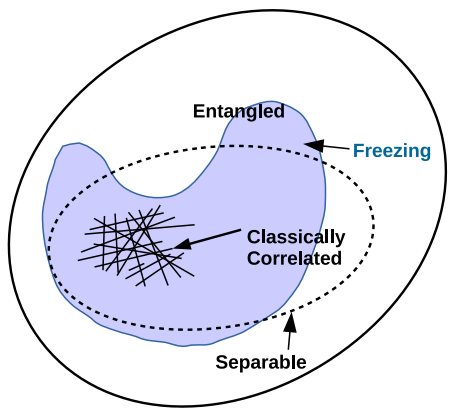}
 \caption{(Color online) States having non-zero quantum correlation that exhibit freezing under local 
 noise are in the shaded region. While the separable states contain the classically correlated states, 
 the freezing states do not, and while the class of separable states is 
 convex, the others are not.}
 \label{schematic}
\end{figure}

\subsection{Relation between entanglement and freezing}

We have already found the conditions by which QD and QWD of two-qubit mixed states 
remain constant with time in the presence of local noise. On the 
other hand, entanglement of the state undergoes sudden death when local BF channels 
are applied~\cite{sd_group}. However, we find that the behavior of entanglement of the CI states of the form 
(\ref{twoqubitstate}) bear interesting correspondence with the freezing behavior of quantum 
correlations. 
For a two-qubit CI state satisfying condition (\ref{necsuf1}), we find that
the region in the $(c_{33},c_{01})$ space, for a fixed $|c_{11}|$, constituting of entangled states, increases 
with increasing $|c_{11}|$ (as shown in Fig. \ref{entphase}), while the freezing regions for QD does the opposite. 
Similar results are found for QWD. Note that an increase (a decrease) of $|c_{11}|$, while satisfying condition (\ref{necsuf1}) 
results in the magnetizations of the state becoming more (less) homogeneous in magnitude.   
For the canonical initial states which satisfy Eq. (\ref{necsuf1}),  
the value of the freezing terminal, $\gamma_{f}$, decreases with increasing $|c_{11}|$. In contrast, entanglement 
lingers for longer time for CI states with high $|c_{11}|$ (i.e., the time at which the entanglement 
becomes zero, increases with the increase of $|c_{11}|$). For a small value of $|c_{11}|$, even separable but 
quantum correlated (as measured by QD or QWD) CI states, when subjected to BF noise, can exhibit 
freezing for a finite interval as exhibited in Fig. \ref{entphase}(a). The dynamics of QD for 
such states with $\gamma$ is depicted in 
Fig. \ref{twoqubitphase}(c). Similar result is obtained for the BD state, in 
Ref.~\cite{mazzola_prl_2010}. The space of 
all mixed states (both separable and entangled) can be classified according to the occurrence and absence of freezing 
of quantum correlations. For a schematic representation, see Fig. \ref{schematic}.

\section{Multipartite freezing states} 
\label{multiqubit}

The question that follows logically from the above discussion is whether freezing is an entirely bipartite 
phenomenon or can also be found in multipartite states. In this section, we demonstrate the freezing of QD 
and QWD in  multiparty systems. For the purpose of demonstration, we use the BF channel. However, similar results 
can be found for other decohering channels also.

\subsection{States with genuine multiparty classical correlators}
Let us consider a quantum state of an even number, 
$2n$, of qubits given by 
\begin{eqnarray}
 \rho_{2n}= \frac{1}{2^{2n}}\left(\otimes_{j=1}^{2n} I_{j}+\sum_{\alpha=1}^{3}c^{\alpha}_{2n}\otimes_{j=1}^{2n}
 \sigma_{j}^{\alpha}\right),
 \label{belllike}
\end{eqnarray}
where $n\geq 1$. We assume $|c_{2n}^{\alpha}|\neq0$. The state is completely defined by the 
\textit{genuine multiparty} \textquotedblleft classical\textquotedblright\; correlators 
$c_{2n}^{\alpha}=\mbox{Tr}\{\left(\sigma^{\alpha}\right)^{\otimes 2n}\rho_{2n}\}$,
where none of the single-qubit operators are multiples of $I$. 
We refer to the state as the \textit{diagonal state}. The marginal states of the above multipartite state in the bipartition 
$j:\mbox{rest} \ (j=1,\cdots,2n)$ are maximally mixed. 
An NS
condition for the freezing of QD, calculated in the partition $j:\mbox{rest} \ (j=1,\cdots,2n)$, for the diagonal state, can be 
obtained using only the genuine multiparty classical correlators.
A similar condition can also be obtained for QWD.  
Note that the two-qubit state that we had
considered before is of rank at most $4$ while the multipartite state here is of rank at most $2^{2n}$.

\noindent\textbf{Proposition VI.} \textit{If local BF noise is applied to a diagonal state,  
an NS condition for freezing of QD in the bipartition where one block consists of a single 
qubit is given by either of the following conditions:} 
\begin{eqnarray}
   \begin{cases}
     (i)  & c_{2n}^{2}=(-1)^{n}c_{2n}^{1}c_{2n}^{3},\\
     (ii) & 1\geq|c_{2n}^{3}|>|c_{2n}^{1}|;\\
   \end{cases}
\label{multifreezingform}
\end{eqnarray}
\begin{eqnarray}
   \begin{cases}
     (i)  & c_{2n}^{3}=(-1)^{n}c_{2n}^{1}c_{2n}^{2},\\
     (ii) & 1\geq|c_{2n}^{2}|>|c_{2n}^{1}|.\\
   \end{cases}
\label{multifreezingform1}
\end{eqnarray}

\noindent\textbf{Proof.} 
Under the application of the BF channel, the time-evolved state $\rho_{2n}^{(\gamma)}$ has the same form as 
that given in Eq. (\ref{belllike}). Both the correlators $c_{2n}^{2}$ and $c_{2n}^{3}$ decay 
with $\gamma$ as $(1-\gamma)^{2n}$ under the BF evolution whereas $c_{2n}^{1}$ remains constant 
over time. For the time evolved state $\rho_{2n}^{(\gamma)}$,  QD in the $j:\mbox{rest}$ bipartition
with $j=1,\cdots,2n$ is given by
\begin{eqnarray}
 D_{2n}=S(\rho_{1})+S(\rho_{2n-1})-S(\rho_{2n}^{(\gamma)})+\frac{1}{2}F(c),
\end{eqnarray} 
where $\rho_{1}$ and $\rho_{2n-1}$ are the reduced density matrices of $\rho_{2n}^{(\gamma)}$, and  
$c=\max\left\{|c_{2n}^{1}|,|c_{2n}^{2}|(1-\gamma)^{2n},|c_{2n}^{3}|(1-\gamma)^{2n}\right\}$. Here, $F(y)$ is the 
freezing entropy defined in Proposition I and $S(\rho_{2n}^{(\gamma)})$, the von Neumann entropy of the state $\rho_{2n}^{(\gamma)}$, can 
be calculated from the eigenvalues of the state $\rho_{2n}^{(\gamma)}$, which are given by 
\begin{eqnarray}
 \lambda_{1}&=&\frac{1}{2^{2n}}(1\pm c_{2n}^{1}\pm c_{2n}^{2}(1-\gamma)^{2n}\pm c_{2n}^{3}(1-\gamma)^{2n}),\nonumber \\
 \lambda_{2}&=&\frac{1}{2^{2n}}(1\pm c_{2n}^{1}\mp c_{2n}^{2}(1-\gamma)^{2n}\mp c_{2n}^{3}(1-\gamma)^{2n}),\nonumber \\
 \lambda_{3}&=&\frac{1}{2^{2n}}(1\mp c_{2n}^{1}\pm c_{2n}^{2}(1-\gamma)^{2n}\mp c_{2n}^{3}(1-\gamma)^{2n}),\nonumber \\
 \lambda_{4}&=&\frac{1}{2^{2n}}(1\mp c_{2n}^{1}\mp c_{2n}^{2}(1-\gamma)^{2n}\pm c_{2n}^{3}(1-\gamma)^{2n}),
\end{eqnarray}
where each of the $\lambda_{i}$  $(i=1,2,3,4)$ are repeated $2^{2n-2}$ times. Note also that the marginal states of 
$\rho_{2n}^{(\gamma)}$ are maximally mixed and are invariant under the local BF evolution.

We first focus on the necessity of the condition (\ref{multifreezingform}). If freezing of QD takes place, $D_{2n}$ must be 
independent of $\gamma$ for a finite interval. Let us first assume that $c=|c_{2n}^{1}|,$ in that interval. Since $c_{2n}^{1}$ remains unaltered under the BF dynamics, and
$\rho_{1}$ and $\rho_{2n-1}$ are independent of $\gamma$, the time dependence in QD comes through the entropy 
$S(\rho_{2n}^{(\gamma)})$. One can easily show that $S(\rho_{2n}^{\gamma})$
varies with $\gamma$ for all possible values of the correlators. This implies that QD does not freeze when $c=|c_{2n}^{1}|$. Next, let us 
take $c=|c_{2n}^{3}|(1-\gamma)^{2n}$. In this case, if QD is frozen over a certain interval of $\gamma$, the correlators must 
satisfy the condition (\ref{multifreezingform})(i) so that the $\gamma$ dependence cancels out and the QD becomes a function of 
$c_{2n}^{1}$ only, thereby proving the necessity of the condition (\ref{multifreezingform}). Similarly, assuming that 
$c=|c_{2n}^{2}|(1-\gamma)^{2n}$, one can prove the necessity of the condition (\ref{multifreezingform1}). 

We now prove the sufficiency of the conditions (\ref{multifreezingform}) and (\ref{multifreezingform1}). Starting with the 
condition (\ref{multifreezingform})(i), one can show that the QD takes the form 
$D_{2n}=\frac{1}{2}(F(c)-F(c_{2n}^{1})-F(c_{2n}^{3}(1-\gamma)^{2n})))$. Application of condition 
(\ref{multifreezingform})(ii) implies that $D=-\frac{1}{2}F(c_{2n}^{1}),$ thereby proving the constancy of the QD for 
a finite time interval. The proof is similar for 
condition (\ref{multifreezingform1}), when the same value of frozen QD is obtained. \hfill $\blacksquare$

Clearly, with the application of condition (\ref{multifreezingform}), freezing sustains 
as long as  $|c_{2n}^{3}|(1-\gamma)^{2n}>|c_{2n}^{1}|$, which gives the value of the freezing terminal, $\gamma_{f}$, as
\begin{eqnarray}
 \gamma_{f}= 1-\left(\frac{|c_{2n}^{1}|}{|c_{2n}^{3}|}\right)^{\frac{1}{2n}},
\end{eqnarray}
with $|c_{2n}^{3}|\neq0$. For fixed $c_{2n}^{1}$, the maximum of $\gamma_{f}$ occurs for $c_{2n}^{3}=\pm1$. Similar 
expression for $\gamma_{f}$ can be obtained from condition (\ref{multifreezingform1}).

\begin{figure}
 \includegraphics[scale=0.65]{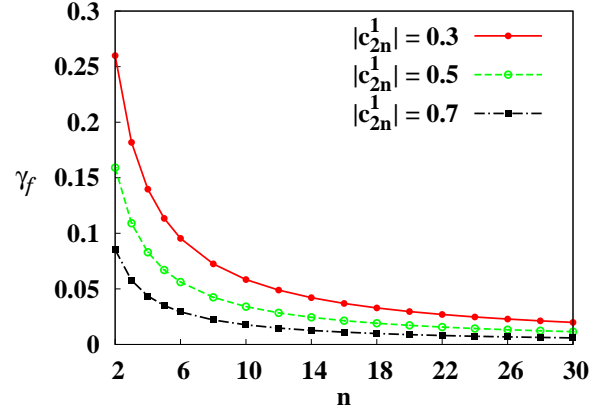}
 \caption{(Color online) Dynamics of the QD in the $1:2\cdots2n$ bipartition for 
 the state given in Eq. (\ref{belllike}) in the case of the BF 
 channel. We plot the variation of $\gamma_{f}$ as a function of $n$ for different values of 
 $\left|c^{1}_{2n}\right|$ with $\left|c^{3}_{2n}\right|=1$, satisfying Eq. (\ref{multifreezingform}).
 All quantities are dimensionless, except the horizontal axis, which is in half of the number of particles.} 
 \label{multi1}
\end{figure}

Note that the value of the frozen discord in the bipartition $j:\mbox{rest}$ is independent 
of the number of parties, $2n$, whereas the 
freezing terminal, $\gamma_{f}$, decreases with increasing $n$, thereby indicating a better freezing with 
low values of $n$, for fixed values of $c_{2n}^{1}$ and $c_{2n}^{3}$. 
Fig. \ref{multi1} depicts the variation of $\gamma_{f}$ with increasing $n$ for different values of $|c_{2n}^{1}|$ 
with $|c_{2n}^{3}|=1$. For fixed values of $n$ and $|c_{2n}^{3}|$, 
$\gamma_{f}$ decreases monotonically with increasing $|c_{2n}^{1}|$ which is also clearly depicted in Fig. \ref{multi1}. 
One should note that it is also possible to incorporate
inhomogeneity in the system by introducing $x$-magnetization in such a way that the magnetization of all the qubits are equal
except for the one over which the measurement is performed in the case of QD and QWD. 
Similar 
results can be derived in the case of the BPF and the PF channels as well. 

\begin{figure}
\includegraphics[scale=0.65]{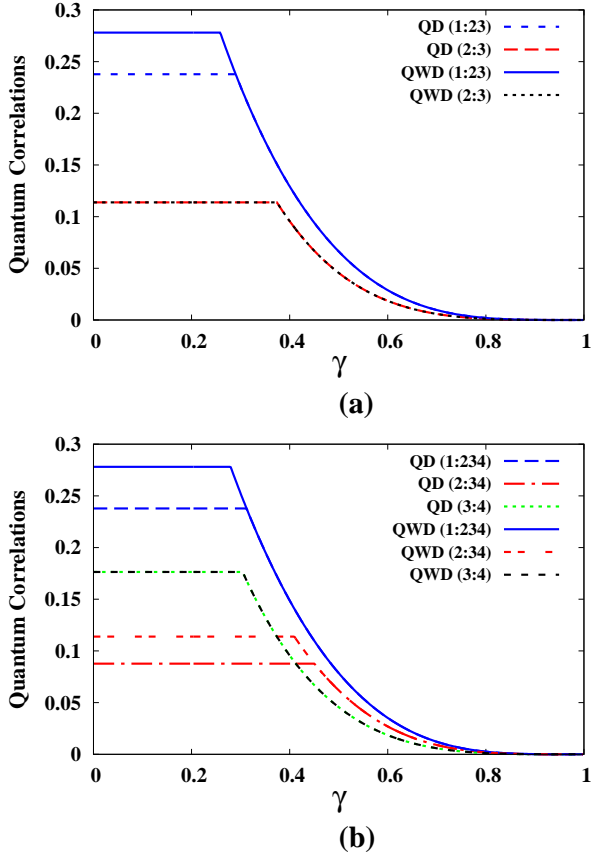}
\caption{(Color online) (a) Freezing of  QD and  QWD for $\rho_{3}(x,\alpha_{1})$ with $\alpha_{1}=0.2$
and $x=0.6$. 
The reduced density matrix $\rho_{3}^{23}(x,\alpha_{1})$ is a BD state for which  QD and  QWD are identical. 
(b) Freezing of quantum correlations for the four-qubit state $\rho_{4}(x,\alpha_{1},\alpha_{2})$ and 
the reduced states $\rho_{4}^{234}$ and $\rho_{4}^{34}$ obtained from 
$\rho_{4}(x,\alpha_{1},\alpha_{2})$ by tracing out the first qubit and the first two qubits respectively. Here, 
$\alpha_{1}=0.2$, $\alpha_{2}=0.25,$ and $x=0.6$. Again, the QD and the QWD are identical for $\rho_{4}^{34}$. All
quantities are dimensionless except QD, which is in bits, and QWD, which is in qubits.}
 \label{sweep}
\end{figure}

\subsection{Sweeping state}
We now propose another prescription for constructing general multiparty freezing states with $n$ qubits, $n$ being even or odd. 
Before presenting the multiparty state, let us first write down an explicit form of the bipartite state
which is a CI state, and which obeys the NS condition (\ref{necsuf1}) with 
$c_{33}^{2}+c_{01}^{2}=1$:
\begin{eqnarray}
 \rho_{2}(x,\alpha)&=&\frac{x}{2}P\left[|\psi^{2}_{0}(\alpha)\rangle+|\psi^{2}_{1}(\alpha)\rangle\right]\nonumber \\
 &&+\frac{1-x}{2}\left[P[|\psi^{2}_{0}(\alpha)\rangle]+P[|\psi^{2}_{1}(\alpha)\rangle]\right],
 \label{qfstateform}
\end{eqnarray}
where $x=c_{11}$, $|\alpha|=\sqrt{\frac{1+c_{33}}{2}}$, 
and $P\left[|\psi\rangle\right]=|\psi\rangle\langle\psi|$.
The states $|\psi^{2}_{0}(\alpha)\rangle$ and $|\psi^{2}_{1}(\alpha)\rangle$ are 
\begin{eqnarray}
|\psi^{2}_{0}(\alpha)\rangle&=&|0\rangle\otimes|\nu^{1}_{0}(\alpha)\rangle, \\
|\psi^{2}_{1}(\alpha)\rangle&=&|1\rangle\otimes|\nu^{1}_{1}(\alpha)\rangle,
\end{eqnarray}
with $|\nu^{1}_{0}(\alpha)\rangle=\alpha|0\rangle+\sqrt{1-\alpha^{2}}|1\rangle$ and 
$|\nu^{1}_{1}(\alpha)\rangle=\alpha|1\rangle+\sqrt{1-\alpha^{2}}|0\rangle$.  
The bipartite state of the form (\ref{qfstateform}) can be straightforwardly extended to the tripartite case as
\begin{eqnarray}
 \rho_{3}(x,\alpha_{1})&=&\frac{x}{2}P\left[|\psi^{3}_{0}(\alpha_{1})\rangle+|\psi^{3}_{1}(\alpha_{1})\rangle\right]\nonumber\\ 
+&&\frac{1-x}{2}\left[P[|\psi^{3}_{0}(\alpha_{1})\rangle]+P[|\psi^{3}_{1}(\alpha_{1})\rangle]\right],
 \label{qfs3}
\end{eqnarray}
with the encoding $|\psi^{3}_{0}(\alpha_{1})\rangle=|0\rangle\otimes|\nu^{2}_{0}(\alpha_{1})\rangle$ and 
$|\psi^{3}_{1}(\alpha_{1})\rangle=|1\rangle\otimes|\nu^{2}_{1}(\alpha_{1})\rangle$, where 
$|\nu^{2}_{0}(\alpha_{1})\rangle=\alpha_{1}|00\rangle+\sqrt{1-\alpha_{1}^{2}}|11\rangle$ and 
$|\nu^{2}_{1}(\alpha_{1})\rangle=\alpha_{1}|11\rangle+\sqrt{1-\alpha_{1}^{2}}|00\rangle$. 
The state in Eq. (\ref{qfs3})
can show freezing of QD as well as that of QWD in the bipartition $1:23$. Note that the marginal state $\rho_{3}^{23}(x,\alpha_{1})=
\mbox{Tr}_{1}\{\rho_{3}(x,\alpha_{1})\}$ is a BD state, which satisfies the freezing condition of QD and QWD, as 
depicted in Fig. \ref{sweep}(a). 

Starting from the state in Eq. (\ref{qfs3}), a four-qubit freezing state $\rho_{4}(x,\alpha_{1},\alpha_{2})$ can be 
generated by  performing a two-qubit encoding in the qubit 3 as   
\begin{eqnarray}
 |0\rangle\rightarrow\nu_{0}^{2}(\alpha_{2})=\alpha_{2}|00\rangle+\sqrt{1-\alpha_{2}^{2}}|11\rangle,\nonumber \\
 |1\rangle\rightarrow\nu_{1}^{2}(\alpha_{2})=\alpha_{2}|11\rangle+\sqrt{1-\alpha_{2}^{2}}|00\rangle.
 \label{encode}
\end{eqnarray}
Freezing of  QD as well as  QWD is observed, when measurement is made on the first qubit of 
$\rho_{4}(x,\alpha_{1},\alpha_{2})$.
Interestingly, like the three-qubit case, all reduced density matrices of $\rho_{4}(x,\alpha_{1},\alpha_{2})$ obtained by tracing 
out parts from left side, starting from the qubit 1, show freezing of QD and QWD. 
In particular, the marginals $\rho_{4}^{234}$, $\rho_{4}^{34}$ of 
$\rho_{4}(x,\alpha_{1},\alpha_{2})$ show freezing of QD and QWD when the bipartition of the marginal state is considered between the
first qubit and the rest of the qubits and the measurements are performed on the first qubit.
The freezing of the four-qubit state and that exhibited by its three- and
two-qubit reduced states are shown in Fig. \ref{sweep}(b). 

The above procedure can be continued to generate an $n$-qubit freezing state $\rho_{n}(x,\{\alpha_{i}\})$, $i=1,\cdots,n-2,$ by 
applying an encoding similar to that in Eq. (\ref{encode}), so that the states $\{|0\rangle,|1\rangle\}$ of the qubit 
$(n-1)$ of $\rho_{n-1}(x,\alpha_{1},\cdots,\alpha_{n-3})$ is now replaced by $\{\nu_{0}^{2}(\alpha_{n-2}),\nu_{1}^{2}(\alpha_{n-2})\}$.
The state $\rho_{n}(x,\{\alpha_{i}\})$ is a very special multipartite state for which 
freezing is observed for QD or  QWD calculated in the bipartition $1:2...n$, 
with the speciality being that a freezing 
state of $m$ parties $(m<n)$ can be obtained from $\rho_{n}(x,\{\alpha_{i}\})$ when $n-m$ parties are traced out from the 
\textquotedblleft left\textquotedblright\; side. 
Each of the $n-m$ states obtained during \textit{sweeping out} the qubits starting from the first qubit is also 
a multipartite freezing state in the bipartition $first \ qubit:rest$, when the freezing is observed by performing the 
measurement on the qubit $(n-m+1)$. We call the state $\rho_{n}(x,\{\alpha_{i}\})$ as the sweeping state.

\section{Effective freezing of quantum correlations: Freezing index}
\label{efreez}

\begin{figure}
 \includegraphics[scale=1.1]{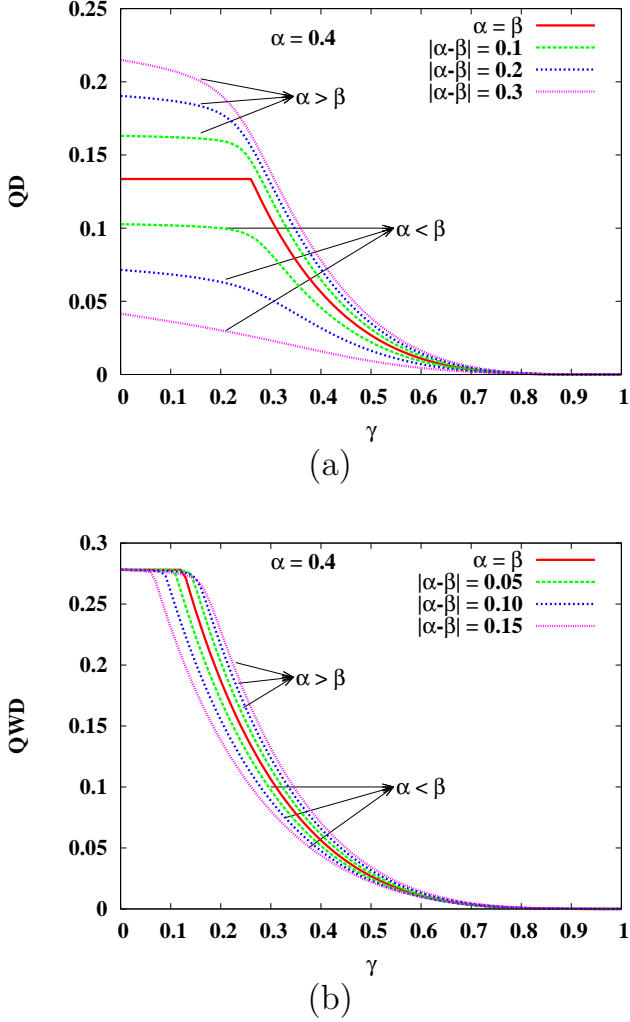}
 \caption{(Color online) The QD (a) and the QWD (b) as functions of the parametrized time $\gamma$, 
 for the initial state given in Eq. (\ref{qfstateform}) in conjunction with Eq. (\ref{alp-bet}) for local 
 BF noise. 
 The value of the parameter $x=0.6$. All quantities are dimensionless except QD, which is in bits, and QWD, which is in qubits.}
 \label{fig-ef}
\end{figure}

There exist classes of bipartite as well as 
multipartite states whose quantum coherence in the form of QD and QWD can remain constant for a finite interval 
of time under a noisy environment. However, such states are special in nature.
Identifying these states are clearly of immense interest for efficient performance of quantum information tasks. From 
a practical viewpoint, it will also be interesting to find states that offer 
very slow decay of quantum correlations, instead of being constant, in time. 
The slowly decaying QD and QWD with time can be termed as \textquotedblleft effective freezing\textquotedblright.
To visualize such phenomena, we plot 
the QD and the QWD, as functions of the parametrized time $\gamma$,  using the initial state given in 
Eq. (\ref{qfstateform}) with
\begin{eqnarray}
 |\psi_{0}^{2}(\alpha)\rangle=|0\rangle\otimes|\nu_{0}^{1}(\alpha)\rangle,\nonumber \\
 |\psi_{1}^{2}(\beta)\rangle=|0\rangle\otimes|\nu_{1}^{1}(\beta)\rangle.
 \label{alp-bet}
\end{eqnarray}
Note that while the QD and the QWD exactly freeze for $\alpha=\beta$, the quantities remain effectively frozen
in a finite time interval, $\Delta\gamma$, for \textquotedblleft small\textquotedblright\; values of $|\alpha-\beta|$, 
as demonstrated in Fig. \ref{fig-ef}.

\medskip 

\noindent\textbf{Freezing index:} 
Let us now introduce a measure, which we call \textquotedblleft freezing index\textquotedblright, to quantify the goodness 
of freezing behavior for a given trio of quantum correlation measure, $Q$, an initial state, and a decoherence channel. 
It necessarily depends on 
\textit{(i)} the value, $Q^{f}$, of the frozen quantum correlation, \textit{(ii)} the duration of freezing, 
$\Delta\gamma^{f}$, \textit{(iii)} the onset of a freezing interval, and \textit{(iv)} the number of freezing 
intervals, $N_{f}$, in the case of the existence of multiple freezing in the dynamics. 
The variation of the quantum correlation measure with respect to time vanishes for 
exact freezing while it is greater than a small number, $\delta$, named tolerance,  for effective freezing.
Note that a given interval is considered to be effectively frozen only if the variation of the quantum correlation measure
at all points in the interval (including the end points) from the value of the measure at the starting point of the interval
remains lower than the tolerance, $\delta$.
In order to quantify the quality of effective freezing, we define a 
\textquotedblleft\textit{freezing index}\textquotedblright, $\eta_{f}$, for an arbitrary quantum correlation measure, as
\begin{eqnarray}
 \eta_{f}=
 \left(\sum_{i=1}^{N_{f}}\overline{Q}^{f}_{i}\left(1-\gamma_{1,i}\right)\int_{\gamma_{1,i}}^{\gamma_{2,i}}Q(\gamma)d\gamma\right)
 ^{\frac{1}{4}},
 \label{fi}
\end{eqnarray}
where $\gamma_{1,i}$ and $\gamma_{2,i}$ are respectively the initial and final points of the ``effective'' freezing interval and
$\overline{Q}^{f}_{i}$ is the average value of $Q$ during the freezing interval.
For both QD and QWD, the maximum value of $\eta_{f}$ is unity, 
which occurs when maximally entangled states are sent through a noiseless channel,  
whereas the minimum value of $\eta_{f}$ is zero. 
Note also that the index can also quantify ``exact'' freezing phenomena, with the ``effective''
freezing interval being replaced by the freezing interval, and $\overline{Q}^{f}_{i}$ being replaced by $Q^{f}_{i}$, the frozen correlation value in the 
freezing interval $i$.

\begin{figure}
 \includegraphics[scale=0.55]{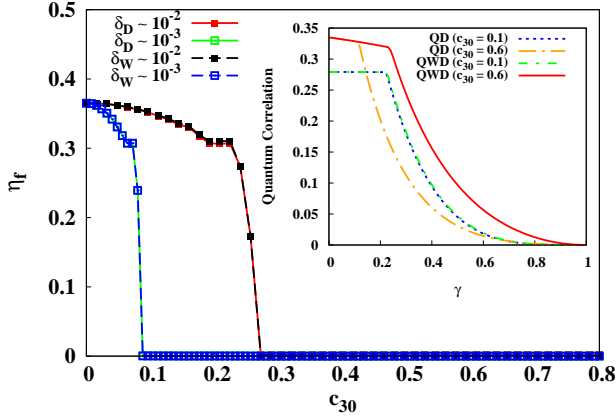}
 \caption{(Color online) Variation of the freezing index against increasing $c_{30}$ in the state in (\ref{efstate}) for 
 QD and QWD, for different values of $\delta$. The suffixes of $\delta$ denote whether QD or QWD is being considered as 
 the measure. We choose $c_{11}=0.6$, $c_{22}=-0.6$, $c_{33}=1.0$, and the local BF channel for the purpose of the plot.
 The curves for $D$ and $W$ for a fixed value of the tolerance merge with wach other.
 Inset: Dynamics of QD and QWD using the state 
 (\ref{efstate}) as the initial state to the BF channel with different values of the magnetization $c_{30}$
 The curves for $c_{30}=0.1$ for QD and QWD have merged with each other. The dimensions are as in Fig. \ref{fig-ef}.}
 \label{efdyn1}
\end{figure}

To demonstrate the freezing index, we consider the bipartite state 
\begin{eqnarray}
  \rho_{AB}&=&\frac{1}{4}[I_{A}\otimes I_{B}+c_{30}\sigma_{A}^{3}\otimes I_{B}+c_{03}I_{A}\otimes\sigma_{B}^{3}\nonumber \\
  && + \sum_{\alpha=1}^{3}c_{\alpha\alpha}\sigma_{A}^{\alpha}\otimes\sigma_{B}^{\alpha}],
 \label{efstate}
\end{eqnarray}
where $\left|c_{30}\right|=\left|c_{03}\right|$, i.e., we have chosen the case of homogeneous magnetization 
in the $z$-direction. 
In general, QD, or QWD is found to be decaying functions 
of time when local BF noise is applied to the state. However, the decay of the QD can be made very slow over a 
certain interval of time, when the state parameters are tuned to appropriate values (see Fig. \ref{efdyn1}). 
For example, for low values of $c_{30}$, with properly chosen other correlators, 
the decay-rates of QD as well as QWD are very low, thereby 
ensuring a high value of $\eta_{f}$. With an increase in the magnitude of $c_{30}$, the effective freezing breaks 
and the correlations decay faster with time. This causes a decrease in the value of $\eta_{f}$.  
The dynamics of QD and QWD with increasing $c_{30}$ is  represented in the inset of Fig. \ref{efdyn1}.

\subsection{Freezing in quantum spin models}

\begin{figure}
 \includegraphics[angle=-90,scale=0.35]{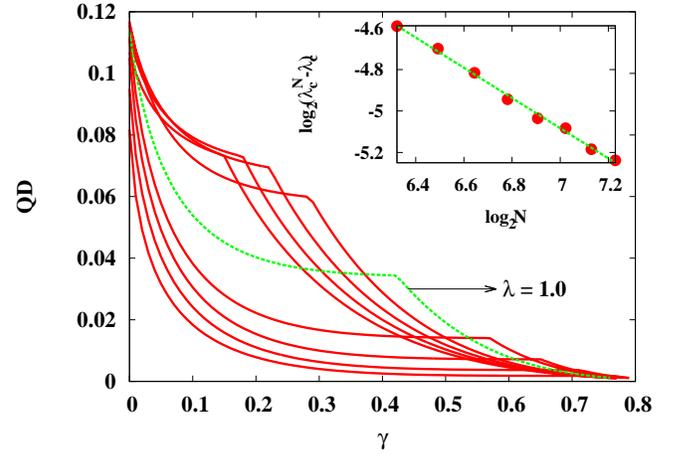}
 \caption{(Color online) QD as function of the parametrized time $\gamma$, for the two-body initial 
 state obtained from the ground state of the infinite spin transverse Ising Hamiltonian (Eq.~\ref{xy}), for 
 values of the parameter $\lambda$ on a equally spaced partition of $[0.6,1.4]$. into $8$ intervals. 
 The curves for $\lambda=0.6$ to $0.9$ lie below the $\lambda=1.0$ curve,
 while those for $\lambda=1.1$ to $1.4$ lie above the $\lambda=1.0$ curve for low values of $\gamma$, say, 
 for $\gamma<0.1$.
 Inset: Finite size scaling analysis for the 1d transverse Ising model using the effective 
 freezing index as the observable. The phase transition point for an $N$ spin system approaches 
 $\lambda=1$ as $N^{-0.729}$. All quantities are dimensionless, except QD, which is in bits, and $\log_{2}N$, which
 is in logarithm of the number of particles.}
 \label{scaling}
\end{figure}

The application of quantum information theoretic concepts and techniques to probe physical phenomena 
in many-body condensed matter systems has given rise to a new cross-disciplinary area of 
research~\cite{modi_rmp_2012,trap_rev,rev_xy}. 
In this section, we investigate the dynamical behavior of the quantum correlation measures when local 
noise is applied to initial states that are ground states of a well-known one-dimensional (1d) 
quantum spin system, namely, the transverse-field anisotropic $XY$ model \cite{xy_group} with periodic 
boundary condition. The Hamiltonian of the model is given by 
\begin{eqnarray}
H_{XY}&=&\frac{J}{2}\sum_{i=1}^{L}
\left\{(1+g)\sigma_{i}^{x}\sigma_{i+1}^{x}+(1-g)\sigma_{i}^{y}\sigma_{i+1}^{y}\right\}\nonumber\\
&&+h\sum_{i=1}^{L}\sigma_{i}^{z}
\label{xy}
\end{eqnarray}
where $J$, $g$ ($-1\leq g\leq1$), and $h$  are respectively the coupling strength, the anisotropy, and the strength of the 
magnetic field.
The model is known to
undergo a quantum phase transition at $\frac{h}{J}\equiv\lambda=\lambda_c\equiv1$ \cite{xy_group,rev_xy,qpt_book}. 
Two special cases of the $XY$ model 
are the transverse-field Ising model  with $g=\pm1$ and the isotropic $XX$ model 
$(g=0)$ in a transverse magnetic field.  
The Hamiltonian $H_{XY}$ can be diagonalized exactly in the thermodynamic limit 
$L\rightarrow\infty$~\cite{xy_group}, for the entire range of values of the anisotropy parameter, 
via the successive applications of the Jordan-Wigner and the Bogoliubov transformations,  
and hence one can determine the nearest- and further-neighbour two-spin reduced density matrices 
for the ground states of the model. 
Since the average transverse magnetization of the ground state in the case of the $XY$ model in a transverse field does not 
vanish, the two-spin states obtained from the ground states do not show exact freezing of QD as well as 
of QWD. We address the issue of effective freezing behavior of quantum correlations in the transverse-field $XY$ model 
and its features in the vicinity of quantum phase transition. We determine the time 
evolved states obtained after local BF channels are applied to the nearest-neighbour density matrices of the ground state.    
In Fig.~\ref{scaling}, we plot the QD as functions of the parametrized time $\gamma$, for a number of two-qubit initial 
states derived from ground states with infinite spins, for different values of $\lambda$ in the vicinity of the quantum 
critical point. The tolerance $\delta$ is fixed at 0.01.
The QD initially decays with time for all values of $\lambda$, after which it effectively freezes for sometime before 
asymptotically decaying to zero. Note that the dynamics of QD at 
the quantum phase transition point stands out from the rest. In particular, an abrupt change in the effective freezing 
index at $\lambda=1$ detects the quantum phase transition. We find that the effective freezing index increases with 
$\lambda$ and vanishes in the paramagnetic region.

The quantum anisotropic $XY$ model with a transverse magnetic field consisting of a finite number of spins 
can be simulated in laboratories~\cite{lab_xy} and therefore, it is important to study the behavior of finite spin systems in 
the context of freezing dynamics. For a finite system, the transition point is again detected by an abrupt change in the 
value of the freezing index. The phase transition point approaches $\lambda_c=1$ with the increase in the size of 
the system as $N^{-0.729}$ i.e.,
\begin{equation}
 \lambda_c^N=\lambda_c+kN^{-0.729},
\end{equation}
where $k$ is a dimensionless constant (see Fig.~\ref{scaling}).

\section{Concluding remarks}
\label{conclude}

In this article, we address an interesting and as yet not entirely understood aspect of the effects 
on the measures of 
quantum correlation belonging to the information theoretic paradigm under decoherence. Specifically, 
we investigate the 
freezing of quantum correlations present in an open quantum system subjected to local 
noise. Our analysis identifies conditions that must be satisfied by 
bipartite as well as multipartite quantum states for freezing of quantum correlations in
a decohering dynamics. It turns out   
that inhomogeneity in the magnetization of the state plays a crucial role in the freezing behavior. 
By comparing freezing properties of QD and QWD, we conclude that the identification of a 
proper measure of correlation is necessary for observation of  freezing in
a specific quantum state, which is clearly in contrast to earlier results. 
We propose a complementarity relation between the frozen value of the quantum correlation and the freezing terminal, 
which is the time at which the quantum correlation in the decohering state ceases to be frozen. 
We also demonstrate the fact that the set of states exhibiting the freezing behavior of quantum correlations is a non-convex set, 
containing entangled as well as separable states.  

We have pointed out that apart from the quantum states that exhibit exact freezing, there also exist many quantum states which 
exhibit extremely slow decay of quantum correlations, and can be appropriate for information theoretic applications. We
 introduce a freezing index -- a quantifier of the figure of merit of the dynamics with respect to freezing, 
which can be useful 
in classifying quantum correlation measures and quantum states with respect to their goodness in freezing. Applying 
the freezing index 
to the transverse-field $XY$ model, we show that the two phases of the ground state of the model have different 
freezing characteristics. The scaling of the freezing index with system-size is also investigated. We expect our approach 
to inspire novel ventures in understanding the intricacies of the dynamics of quantum correlations in open 
quantum systems. 

\acknowledgments

We thank A. Bhattacharya and M. Masud for useful discussions. 
We acknowledge computations performed at the cluster computing facility of
Harish-Chandra Research Institute.

\end{document}